\documentclass[prd, twocolumn, nofootinbib, floatfix]{revtex4-1}

\usepackage{amsmath}
\usepackage{graphicx}
\usepackage{dcolumn}
\usepackage{bm}
\usepackage{epsfig}
\usepackage{amssymb,latexsym,mathrsfs}
\usepackage{graphicx}
\usepackage{color}
\usepackage{hyperref}

\usepackage{amsmath,amssymb}
\usepackage{hyperref}
\usepackage{subfigure}
\usepackage{setspace}
\usepackage{multirow}

\hypersetup{
    colorlinks=true,
    linkcolor=red,
    citecolor=blue,
}

\def\be{\begin{equation}}
\def\ee{\end{equation}}
\def\ba{\begin{eqnarray}}
\def\ea{\end{eqnarray}}
\def\bea{\begin{eqnarray}}
\def\eea{\end{eqnarray}}
\def\nn{\nonumber}

\def\reff{Eq.~(}
\def\refff{Eqs.~(}

\def\be{\begin{equation}}
\def\ee{\end{equation}}
\def\ba{\begin{eqnarray}}
\def\ea{\end{eqnarray}}
\def\nn{\nonumber}

\begin{document}

\title{Degeneracies in parametrized modified gravity models}

\date{\today}

\author{Alireza Hojjati}
\vspace{0.5cm}

\affiliation{Institute for the Early Universe,
Ewha Womans University, Seoul, 120-750, South Korea\\
Department of Physics, Simon Fraser University, Burnaby, BC, V5A 1S6, Canada }

\begin{abstract}

We study degeneracies between parameters in some of the widely used parametrized modified gravity models. We investigate how different observables from a future photometric weak lensing survey such as LSST, correlate the effects of these parameters and to what extent the degeneracies are broken. We also study the impact of other degenerate effects, namely massive neutrinos and some of the weak lensing systematics, on the correlations.  

\end{abstract}

\maketitle

\section{Introduction}

The idea of modified gravity (MG) has been pursued as an alternative to dark energy (DE) in explaining the observed cosmic acceleration. 
Most of the proposed DE/MG theories predict more or less the same background evolution for the universe so that distance measurement data, such as CMB and Supernovae (SNe), can not effectively discriminate between them. On the other hand, these theories have different predictions for the evolution of structure \cite{Koyama:2005kd,Song:2006ej,Pogosian:2007sw,Dvali:2007kt,Silvestri:2009hh} that can be measured to a high precision by the ongoing and upcoming large-scale structure surveys \cite{Zhao:2009fn,Zhao:2008bn,Song:2010fg, Giannantonio:2009gi,Daniel:2010ky,Bean:2010zq}.

The growth of structure can be tested in a model-independent way by parametrizing the evolution equations of the cosmological perturbations.
At the linear level, it is achieved by introducing two scale- and time-dependent functions (MG functions) in the equations relating the scalar metric perturbations, $\Phi$ and $\Psi$, to each other and to the matter perturbations \cite{Baker:2012zs,Amendola:2007rr,Bertschinger:2008zb,Zhao:2008bn,Zhao:2009fn,
Pogosian:2010tj,Caldwell:2007cw,Jain:2007yk,Kunz:2006ca,Wei:2008vw}.  

One of the goals of taking such a phenomenological approach is to detect possible deviations of perturbations from their default evolution,  predicted by $\Lambda$CDM +GR, in an efficient and model-independent way. Therefore, it is important to study the degeneracies between the parameters in these models and their impact on the parameter constraints. One can investigate the scales and redshifts where these degeneracies are minimum, how different observables can break these degeneracies, or what combinations of data do it best.  This can help to come up with an optimum set of parameters for extracting information from data and constraining the MG parameters. 

In this paper we work with two sets of MG functions. The first set parametrizes the modified Poisson equation, $\mu(k,a)$, and the ratio between the two Newtonian potentials, $\gamma(k,a)$ \cite{Zhao:2009fn,Baker:2012zs,Bertschinger:2008zb}. In the second case, we use $\mu(k,a)$ and also parametrize the modifications to the weak lensing (WL) potential by $\Sigma(k,a)$ \cite{Pogosian:2010tj,Song:2010rm,Baker:2012zs}. We consider the approximate parametric form of these functions for some of the widely used MG theories in the literature and quantify how different data sets can improve on degeneracy breaking of the parameters. We use CMB temperature and polarization data from the Planck satellite \cite{Planck}, SNe from an Euclid-like survey \cite{euclidsn} and large scale structure data from a photometric WL survey such as {Large Synoptic Survey Telescope} (LSST)~\cite{LSST}. We also consider pixelations of ($\mu,\gamma$) as a more model-independent treatment of these functions. 
With the pixels of $\mu$ and $\gamma$, one is able to study the degeneracies at particular redshifts and scales of interest.

There are different physical processes that lead to degenerate effects with modified gravity on the cosmological observables. For example, massive neutrinos can modify the growth of structure on their free-streaming scale. This modification can, in principle, be degenerate with the effects of MG on the overlapping scales and redshifts \cite{Abazajian:2011dt,Saito:2010pw,Thomas:2009ae}. We also consider the impact of some of the systematic effects expected in WL surveys. We use \cite{Huterer:2005ez,Zhan:2008jh} where three sources of systematics are modeled for future tomographic surveys: photo-$z$ errors, additive and multiplicative errors due to the uncertainty of the point spread function (PSF) measurements.

In Section \ref{Formalism}, we describe the formalism, MG models, observables and experiments we use. In Section \ref{Degeneracy}, we study the degeneracies between the MG and other parameters of our chosen models where different parametric functional forms of the MG functions are used. From a combination of the upcoming CMB, SNe, galaxy number counts (GC) and WL data, we show that parametrizations with small number of MG parameters can not capture a wide range of possible MG features in the data while too many parameters mainly introduces degeneracies with no extra benefit.
In Section \ref{additional}, we investigate how the additional effects (mass of neutrinos and WL systematics) will change the results. The inclusion of such degenerate effects would require other complementary data to reduce the correlations between the parameters. We summarize in Section \ref{summary}.

\section{The Formalism}
\label{Formalism}

\subsection{Parametrization of the linear perturbations}

The line element for a linearly perturbed FRW universe can be written in the Newtonian gauge as 
\begin{equation}\label{FRW}
ds^2=-(1+2\Psi)~d\tau^2+a^2(t)~(1-2\Phi)~d\vec{x}^2\,,
\end{equation}
where $\Psi$ and $\Phi$ are the scalar perturbations of the metric, $\tau$ the conformal time and $x$ the spatial coordinate. 

At the level of linear perturbations, in addition to $\Psi$ and $\Phi$, one needs to evolve the matter density contrast $\delta\equiv \delta\rho/\rho$ and the velocity field perturbation $v$. Two equations are obtained from the conservation of energy momentum tensor $T^{\mu \nu}_{; \mu} = 0$ \cite{Ma:1995ey}
\begin{eqnarray}
	\dot{\delta} &=& - (1+w) \left(\theta+{\dot{h}\over 2}\right)
	  - 3{\dot{a}\over a} \left({\delta P \over \delta\rho} - w
	  \right)\delta  \,,  \label{deltadot}\\
\dot{\theta} &=& - {\dot{a}\over a} (1-3w)\theta - {\dot{w}\over
	     1+w}\theta + {\delta P/\delta\rho \over 1+w}\,k^2\delta
	     - k^2 \sigma \label{vdot} \,,
\end{eqnarray}
where $\theta \equiv k v$, $w$ is the equation of state parameter for each fluid component and $\sigma$ is the anisotropic stress term defined as $(\rho+P)\sigma \equiv -(\hat{k}_i \hat{k}_j-\frac{1}{3}\delta_{ij})\pi_{ij}$ with $\pi_{ij}$ being the traceless component of the energy-momentum tensor.

The other two equations required to close the system are provided by the gravity theory of the model. In the case of general relativity (GR), the equations are 
\ba
&&k^2\Phi = - 4 \pi G a^2 \sum \rho \Delta \label{poisson-GR} \, , \\
&&k^2(\Phi - \Psi) = 12 \pi G a^2   \sum(\rho + P) \sigma \label{anisotropy-GR} \, ,
\ea
where $\rho \Delta = \rho \delta + 3 (aH/k) v$ is the comoving density perturbation and the sum is over all the species present in the universe. 

In the parametrized framework, \refff\ref{poisson-GR}) and (\ref{anisotropy-GR}), are modified by introducing the MG functions to accommodate possible deviations from GR (+$\Lambda$CDM) 
\ba
&&k^2\Psi = - \mu(k,a) 4 \pi G a^2 \sum \lbrace \rho \Delta + 3(\rho + P) \sigma \rbrace  \label{Poisson} \, , \\
&&k^2[\Phi - \gamma(k,a) \Psi] = \mu(k,a)  12 \pi G a^2  \sum (\rho + P) \sigma \label{anisotropy} \,  ,
\ea
where at late times the shear term, $\sigma$, is negligible and cold dark matter (CDM) perturbations are dominant so that the above equations are simplified. The reason for using the scalar potential $\Psi$ in \reff\ref{Poisson}), as opposed to $\Phi$, is that the growth of matter inhomogeneities is more directly related to $\Psi$. To see this note that the growth of matter on sub-horizon scales can be derived by combining \refff\ref{deltadot}) and (\ref{vdot}) in the limit of $k/\mathcal{H} \rightarrow \infty$ 
\be 
\label{matter-sub-evolution}
\ddot{\delta} + \mathcal{H} \dot{\delta} + k^2 \Psi =0\,,
\ee
where $\mathcal{H} \equiv \frac{\dot{a}}{a}$ and time derivative is w.r.t. conformal time. 

In this approach, $\mu(a,k)$ and $\gamma(a,k)$ are generic functions that parametrize {\it solutions} in alternative gravity theories and depend on the choice of the initial conditions. In some theories, one can calculate their approximate form analytically in the quasi-static approximation. For instance, perturbation equations have been derived for some classes of scalar-tensor theories in \cite{Bertschinger:2008zb,Zhao:2008bn,Brax:2005ew,Song:2006ej,Bean:2006up,Pogosian:2007sw,Tsujikawa:2007xu}, and for the DGP theory  and its higher-dimensional extensions in ~\cite{Dvali:2000hr,Koyama:2005kd,Song:2006jk,Song:2007wd,Cardoso:2007xc,Afshordi:2008rd}.

\subsection{Experiments and observables}

The data considered here include CMB temperature and polarization (T and E), SNe observations, weak lensing (WL) shear of distant galaxies, galaxy number counts (GC), and their cross-correlations.

\subsubsection{CMB and SNe Ia data}
\label{sec:cmbsn}

We assume the expected CMB data from the Planck mission~\cite{Planck} of the {European Space Agency} (ESA) using the same parameters as in \cite{Zhao:2008bn, Hojjati:2011xd}. In addition, to better constrain the background expansion parameters, we include simulated SNe luminosity data for a Euclid-like survey \cite{euclidsn}. We generate 4012 data points randomly distributed in 14 redshift bins from $z = 0.15$ to $z = 1.55$. The nuisance parameter ${\cal M}$, is treated as an undetermined parameter in the analysis.

\subsubsection{Large scale structure data}

We assume the GC and WL data by the {Large Synoptic Survey Telescope} (LSST)~\cite{LSST}. LSST is a proposed large aperture, ground-based, wide field survey telescope. It is expected to cover up to half of the sky and
catalogue several billion galaxies out to redshift $z\sim 3$. For LSST, we adopt parameters from the recent review paper by the LSST \cite{Ivezic:2008fe}.

\subsubsection{Observables}

We consider observables from the above surveys and use the angular power spectra from auto-correlation and cross-correlation of these observables for our Fisher analysis. The power spectra are calculated using MGCAMB code \cite{Hojjati:2011ix,Zhao:2008bn} where \refff\ref{Poisson}) and (\ref{anisotropy}) are implemented. With LSST, we consider 10 and 6 tomographic redshift bins for GC and WL observations, respectively. 

The details of the implementations are described in \cite{Hojjati:2011xd}. 

\begin{table*}[t]
\begin{tabular}{|c|c|c|c|c|}
\hline
Parameter description  & I & II & III &  IV         \\
\hline
$\Omega_bh^2$, baryon density 			& 1  & 1  & 1 &  1\\
$\Omega_ch^2$, cold dark matter density & 2  & 2  & 2 &  2 \\
$H_0$, Hubble constant today 			& 3& 3& 3 &3\\
$\tau$, optical depth 					& 4& 4& 4 &4\\
$n_s$, scalar spectral index 			& 5& 5& 5 &5\\
$w_i$	,effective DE equation of state & 6& 6& 6 & 6-7\\
$MG$, MG parameters in each model			& 7& 7-9& 7-11 & 8-19\\	
$A_s$, amplitude of scalar perturbations& 8& 10& 12 & 20\\
$b_i$, bias parameters for 10 GC tomographic redshift bins & 9-18&11-20&13-22&21-30\\
 \hline
\end{tabular}
\caption{List of the parameters included in our models (I - IV), with the numbers showing the order of their appearance in each model.} 
\label{table:params}
\end{table*}

\subsection{MG models}
\label{section:models}

We choose to work with four models which include one, three, five and twelve MG parameters, respectively. Table \ref{table:params} shows the model parameters that we use in our Fisher analysis and the order in which they are sorted. In all of the models, we assume spatially flat geometry and include the main cosmological parameters ($\#$1-5 and $A_s$). When GC data is included, we assume that the bias is scale-independent on the linear scales and introduce 10 constant bias parameters, one for each photometric redshift bin surveyed by LSST. $w(z)$ is either binned or considered as a constant and is included in the Fisher analysis \footnote{We include the SNe Ia nuisance parameter in all these models and then marginalize over it.}.

\subsubsection{1-parameter MG model: f(R) theory}

$f(R)$ models can be tuned to reproduce any background expansion history, and the remaining relevant quantity is the squared Compton wavelength of the new scalar degree of freedom $f_R \equiv df /dR$ mediating the fifth force. In units of the Hubble length squared it is given by \cite{Hu:2007nk,Song:2006ej}
\begin{eqnarray}
\label{B}
B \equiv {f_{RR} \over 1+f_R} {d R \over d\ln a} \left(  d\ln H \over d\ln a \right)^{-1}\,,
\end{eqnarray}
where $R$ is the background Ricci scalar. Thus, for a fixed background expansion history, different $f(R)$ models can be parametrized by the parameter $B_0$, which is the value of $B$ today. It was suggested in \cite{Giannantonio:2009gi} that for $B_0 \ll 1$ the large scale growth in $f(R)$ models can be modeled as: 
\be\label{eq:par_mu2}
\mu(k,a)=\frac{1+\frac{4}{3} B_0 \mathcal{C}}{1+ B_0 \mathcal{C}} \, \quad \gamma(a,k)=\frac{1+\frac{2}{3} B_0 \mathcal{C}}{1+\frac{4}{3} B_0 \mathcal{C}}\,,
\ee
where $\mathcal{C} \equiv (c k a^2)^2/2H_0^2$. This one-parameter model gives a good approximation of $f(R)$ theories in the quasi-static limit~\cite{Pogosian:2007sw,Giannantonio:2009gi}. We choose a fiducial value of $B_0 \sim 10^{-6}$ in accordance with the forecasted error on this parameter from LSST \cite{Hojjati:2011xd}.

\subsubsection{3-parameter MG model: scalar-tensor theory}

An alternative parametrization to \refff\ref{Poisson},\ref{anisotropy}) can be written as \cite{Song:2010fg,Pogosian:2010tj}
\ba
\label{musigma_mu}k^2\Psi &=& -4\pi G a^2\mu(k, a) \rho \Delta ,\\
\label{musigma_sigma}k^2 (\Phi + \Psi) &=& -8\pi G a^2 \Sigma(k, a) \rho \Delta ,
\ea
where the lower equation governs the perturbations of the relativistic particles through light deflection and ISW effect, and $\Sigma(k, a)$ accounts for deviations from the default evolution in the $\Lambda$CDM+GR model. It has been shown that the following choice can cover a wide range of scalar-tensor theories available \cite{Zhao:2011te}:

\begin{eqnarray}
\mu=1+\frac{c a^s (k/H_0)^n}{1+3c a^s (k/H_0)^n} \,, \quad \Sigma \approx 1 \,. \label{musigma}
\end{eqnarray}
It is easy to show that \refff\ref{musigma}) can be described in terms of ($\mu,\gamma$) as
\begin{eqnarray}
\mu&=&1+\frac{c a^s (k/H_0)^n}{1+3c a^s (k/H_0)^n} \,, \nn \\ 
\gamma \equiv \frac{2}{\mu} -1 &=& 1 - \frac{6 c a^s (k/H_0)^n}{1+c a^s (k/H_0)^n} \,. \label{musigmagamma}
\end{eqnarray}
We work with the fiducial values of $s=4$ and $n=2$ and $c = 0.002$. The value for $c$ is taken from the $ 95\%$ C.L. constraints on the parameter from the current large scale structure data \cite{Zhao:2011te}.

\subsubsection{5-parameter MG model: a general BZ parametrization}

For a wide class of scalar-tensor theories, ($\mu,\gamma$) can be described by the following approximate expressions (BZ parametrization) \cite{Zhao:2009fn,Bertschinger:2008zb}

\begin{eqnarray}
&&\mu(a,k)=\frac{1+\beta_1\lambda_1^2\,k^2a^s}{1+\lambda_1^2\,k^2a^s} \,, \nn \\
\label{BZ}
&&\gamma(a,k)=\frac{1+\beta_2\lambda_2^2\,k^2a^s}{1+\lambda_2^2\,k^2a^s} \ .
\label{BZ}
\end{eqnarray}
where $\beta_1$, $\beta_2$, $\lambda_1$, $\lambda_2$ and $s$ can have different values in different MG theories. The BZ parametrization  provides more flexibility in capturing MG features. We choose a set of fiducial parameter values corresponding to a f(R)-class model \cite{Zhao:2008bn}: $\beta_1 = 4/3$, $\beta_2 = 1/2, s =4, \lambda_s^2 = \beta_1 \lambda_1^2 = 1000$ Mpc$^2$.

\subsubsection{12-parameter MG model: model-independent pixelization}

We also consider the less model-dependent case where $\mu$ and $\gamma$ are "pixelized" in $z$ and $k$ space. We choose to have three bins in the linear $k$ range ($10^{-5}-0.2$ h/Mpc) and for each $k$ bin, we have a \textit{high-z} bin ($3 < z < 30$) and a \textit{low-z} bin ($0 < z < 3 $). In total, there are twelve MG pixels, six for $\mu$ ($\#$ 8-13) and six for $\gamma$ ($\#$ 14-19). $w(z)$ is also binned to one low-z bin ($\#$6) and one high-z one ($\#$7).

\section{Degeneracy of Parameters}
\label{Degeneracy}

In this section, we study the correlations between the MG and other parameters of the models described in \ref{section:models}. For each MG parameter, we show two correlation plots:
\begin{itemize}
\item \textit{Un-normalized} correlation $C_{ij}$, where $C$ is the covariance matrix of all parameters, computed by inverting the  corresponding Fisher matrix.
\item \textit{normalized} correlation: $\overline{C}_{ij} \equiv C_{ij}/\sqrt{(C_{ii}\times C_{jj})}$.
\end{itemize}  
The reason for showing the un-normalized correlations is that it is hard to judge about the importance of correlation between two parameters by just looking at the normalized correlation values. 
For example, a parameter might be highly correlated with the others while being very well-constrained, which means that it is well-measured anyway. For the un-normalized case, we plot the absolute values of correlations on a semi-log plot. Also we show the $\sqrt{C_{ii} \times C_{jj}}$ values where $i$ runs over all the parameters and $j$ is the index of the MG parameter we are plotting the correlations for. These values will be used as a reference and help us estimate the relative degree of correlations between parameters. 

We start from the combination of the CMB and SNe Ia data. We then add the GC and WL data in two steps and study how including them breaks the degeneracies. Finally, we add all the cross-correlations of CMB$\times$CG, CMB$\times$WL and WL$\times$CG. As mentioned, CMB+SNe data are always included and mainly constrain the background parameters. For simplicity, we do not show CMB+SNe and CMB+SNe+WL combinations in the un-normalized plot.

With CMB+SNe data only, one measures the cosmological distances and thus, the background parameters. To describe this, we note  that the comoving distance to a redshift $z$ is defined as  
\be  
\chi \propto \int_{0}^{z} \frac{dz}{H_0 \left[\Omega_m (1+z)^{3} + \Omega_r (1+z)^{4} + \Omega_{DE}(z)\right]^{1/2} } ,
\ee 
where $\Omega_m$, $\Omega_r$ and $\Omega_{DE}(z)$ are the matter, radiation and dark energy fractional densities.
Varying the background parameters changes the comoving distances to observed phenomena such as recombination. This interferes with the growth of perturbations by changing the time available for perturbations to grow, which can be degenerate with changing the growth rate.

The growth of the matter inhomogeneities is probed by large scale structure surveys through estimation of the scalar potential $\Psi$ in \reff\ref{matter-sub-evolution}). With the GC data, one can measure the density fluctuations of galaxies ($\rho_g \delta_g$) which is related to the total matter density fluctuations, $\rho_m \delta_m$, via the bias parameter
\be 
\delta_g = b ~ \delta_m \ .
\ee
In other words, the potential $\Psi$ is estimated from the matter density fluctuations via \reff\ref{Poisson}) up to a (linear) bias factor. 
Since baryons fall into the same gravitational potential as CDM, we can write
\be 
\label{matter_growth-GC}
\ddot{\delta_g} + \mathcal{H} \dot{\delta_g} \propto \frac{\mu}{b}  ~ \delta_g \,.
\ee
Having $\delta_g$ measured by GC observations, the growth of matter perturbations is directly related to $\mu$, bias and the Hubble parameter. This demonstrates the source of large degeneracies between the MG parameters, bias and the background parameters.

\begin{figure*}[th!]
\begin{center}
\includegraphics[scale=0.5]{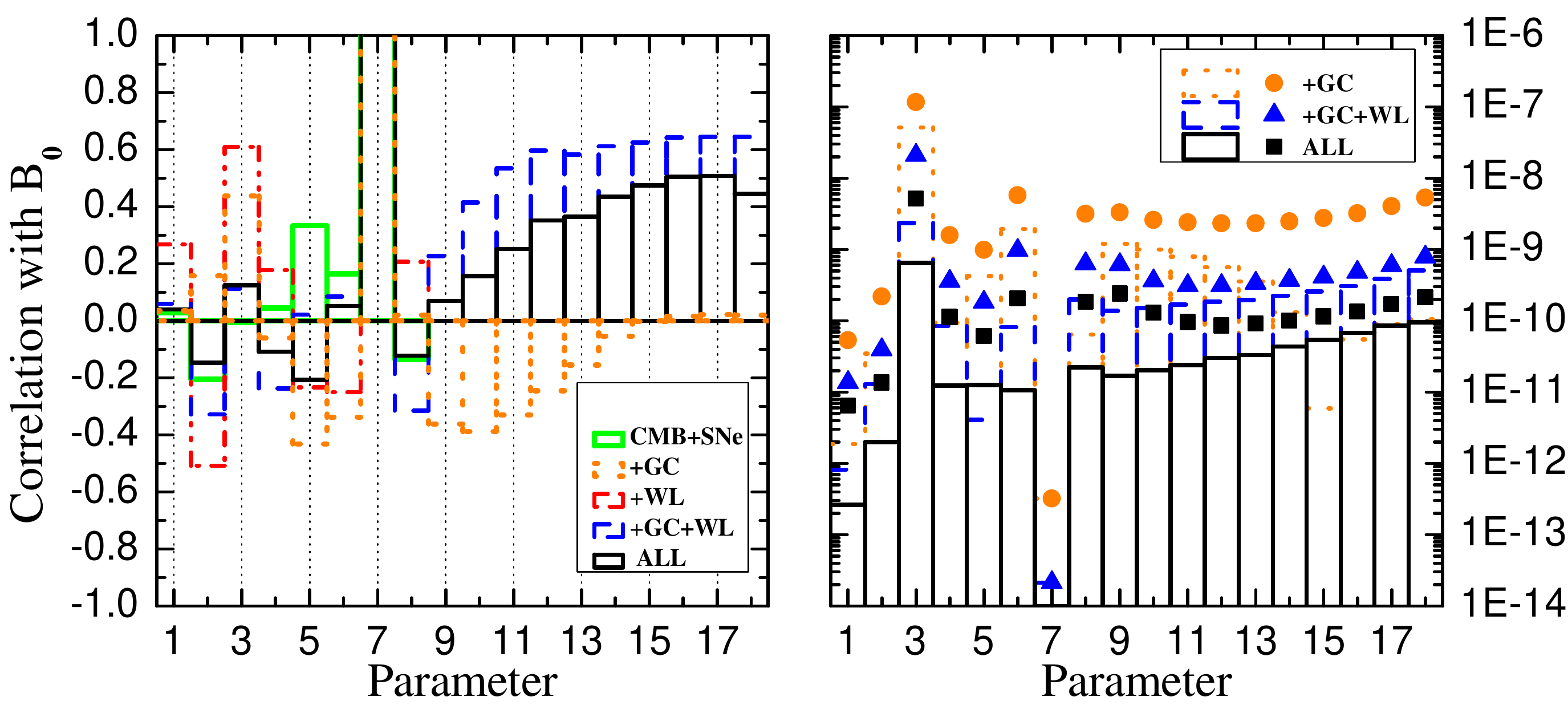}
\caption{Correlation between $B_0$ ($\# 7$) and other parameters for different combinations of data, normalized (left) and un-normalized (right).}
\label{fig:LSST_B0_wCDM}
\end{center}
\end{figure*}

\begin{figure*}[t]
\centering
\subfigure
{
\includegraphics[width=2\columnwidth]{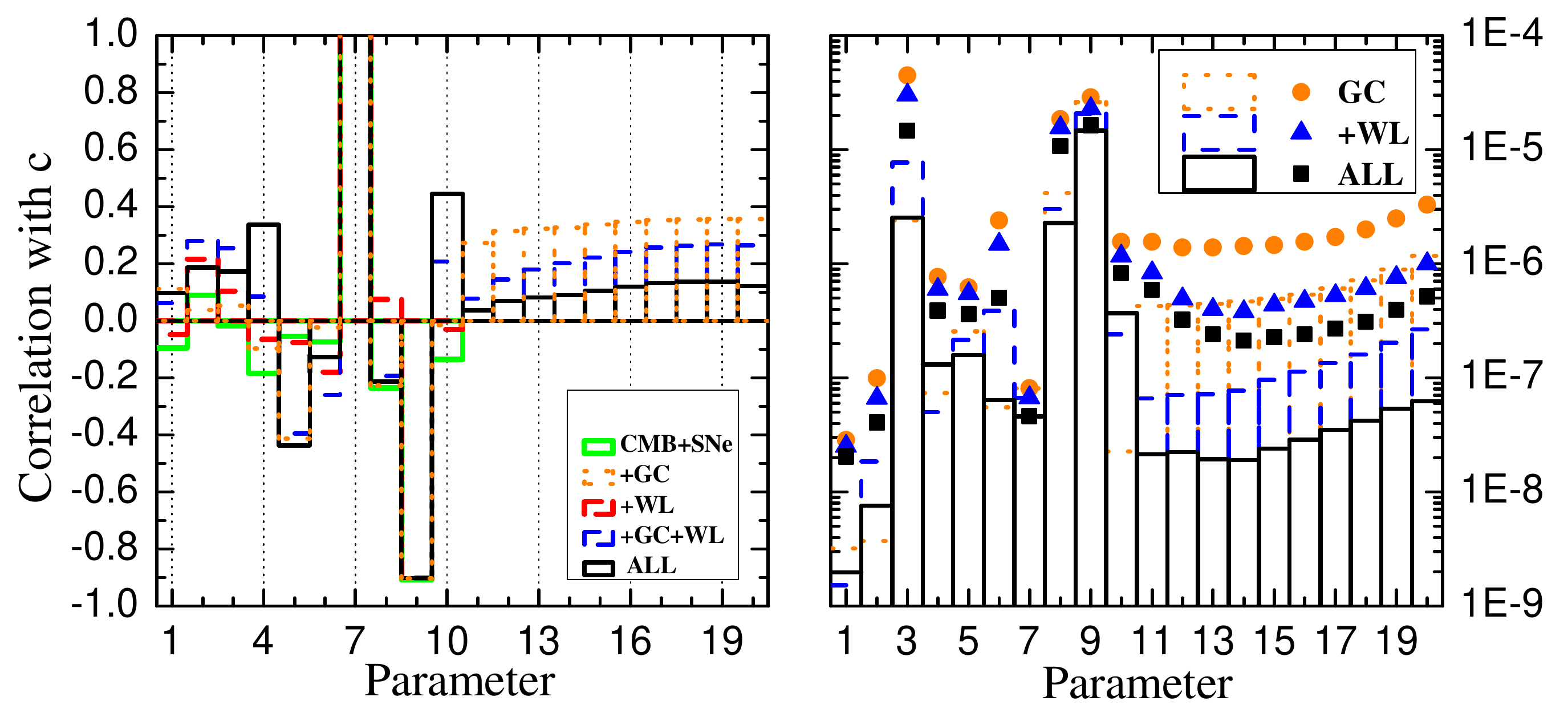}
\label{fig:LSST_csn_wCDM-C}
}
\subfigure
{
\includegraphics[width=2\columnwidth]{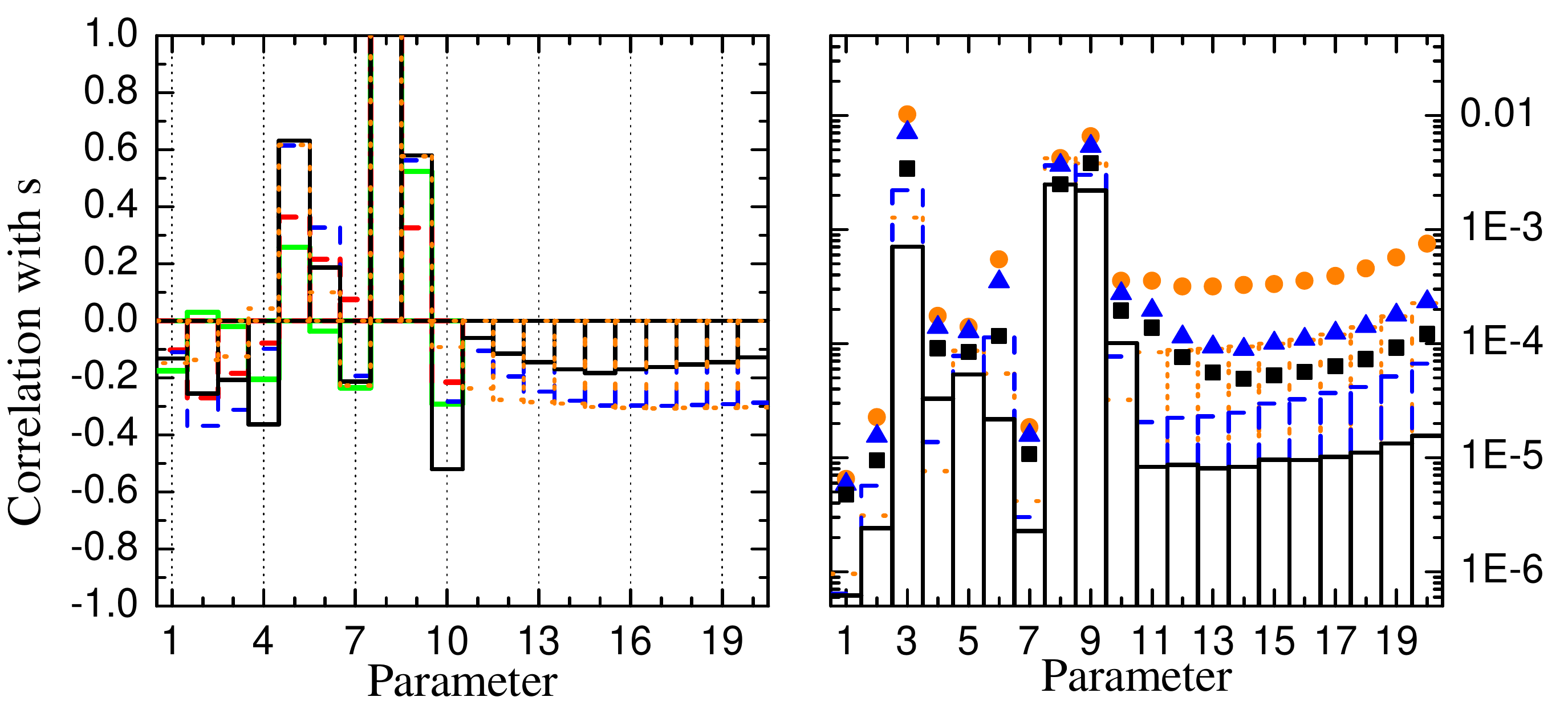}
\label{fig:LSST_csn_wCDM-s}
}
\subfigure
{
\includegraphics[width=2\columnwidth]{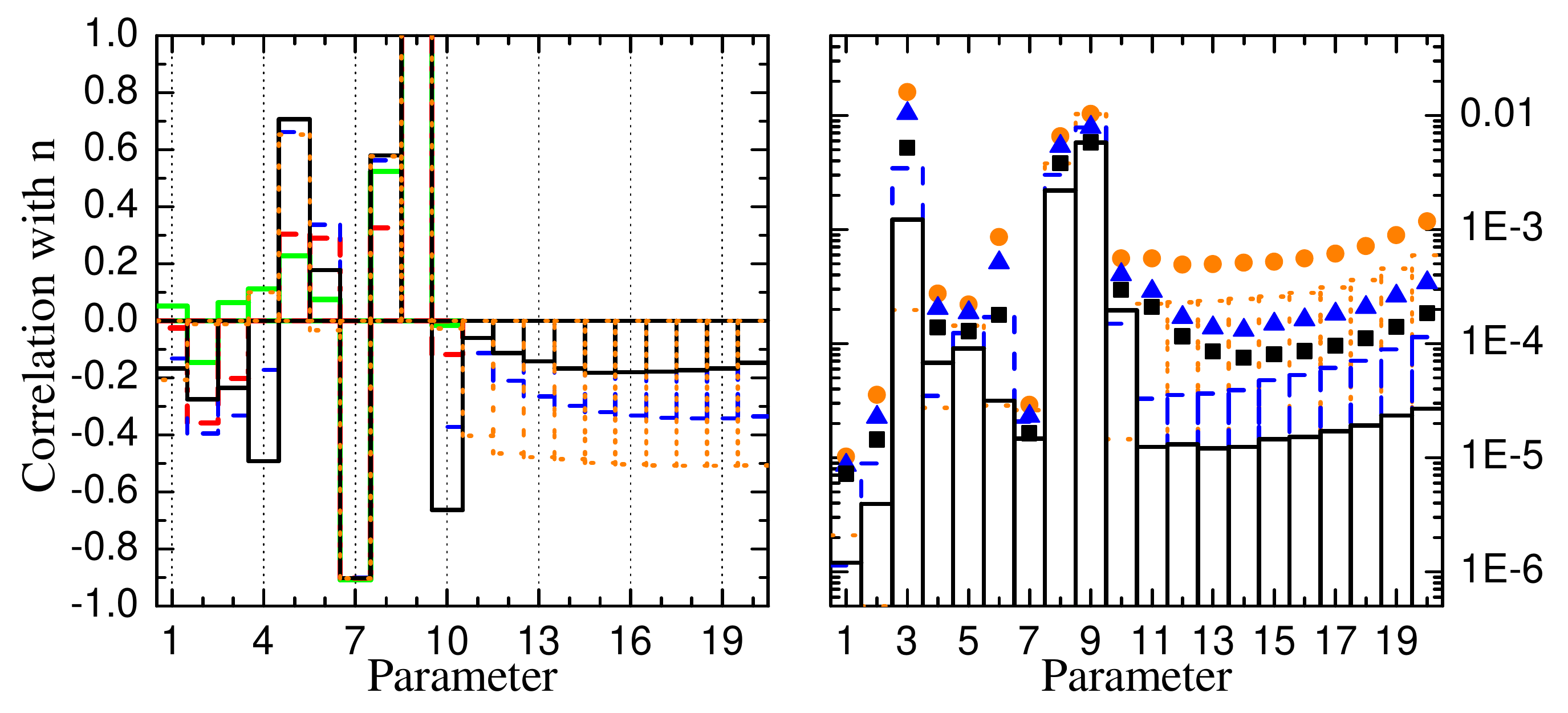}
\label{fig:LSST_csn_wCDM-n}
}
\caption{Correlation between $c$ (top row), $s$ (middle row) and $n$ (bottom row) and other parameters for different combinations of data, same as in Fig.~\ref{fig:LSST_B0_wCDM}. }
\label{fig:LSST_csn_wCDM}
\end{figure*}

WL data probes the $\Phi+\Psi$ combination. With a similar argument as above, one can see that the evolution of the WL potential is determined by $\mu$ and $\Sigma$ (\refff\ref{musigma_mu}) and (\ref{musigma_sigma})) and hence, a combination of $\mu$ and $\gamma$.
Here, the bias parameter is not involved and the growth of matter is directly dependent on $\rho_m \delta_m$. With WL data alone, there is a high degeneracy between the MG parameters and $\Omega_b h^2$, $\Omega_c h^2$ and $H_0$ (Figs.~\ref{fig:LSST_B0_wCDM}, \ref{fig:cont-B0}, \ref{fig:LSST_csn_wCDM}, \ref{fig:LSST_BZ_wCDM}, \ref{fig:LSST_6-6-2-corr}).

When both GC and WL data are included, different combinations of the potentials $\Phi$ and $\Psi$ are probed from independent measurements of $\rho_g \delta_g$ and $\rho_m \delta_m$. It helps break the degeneracies (in particular with the bias parameters) significantly. When adding the cross-correlations of GC and WL, further information about the time evolution of the perturbations at multiple redshift bins is provided. In fact, as we will see, the cross-correlation data puts the strongest constraints on the MG parameters and breaks the degeneracies best.

\subsection{f(R) models parametrized by $B_0$}

$B_0$ is related to the Compton wavelenght of the scalar degree of freedom (DoF). Below this transition scale, the growth of structure is modified due to the presense of the fifth force. It should be kept in mind when one wants to explain the correlations of $B_0$ with other parameters.  Fig.~\ref{fig:LSST_B0_wCDM} shows the normalized (left) and un-normalized (right) correlations between $B_0$ ($\# 7$) and the rest of the model parameters.

As expected, the constraints (shown by scattered plots) are tighter and the correlations (the bar plots) are decreasing when new data included. With all data, except for the bias parameters, the correlations (black bars) are about $\sim$ 10$\%$ or less when compared to the parameter constraints (the $\sqrt{C_{B_0B_0} \times C_{jj}}$ values shown with black squares).

\begin{figure*}[t!]
\centering \includegraphics[scale=0.45]{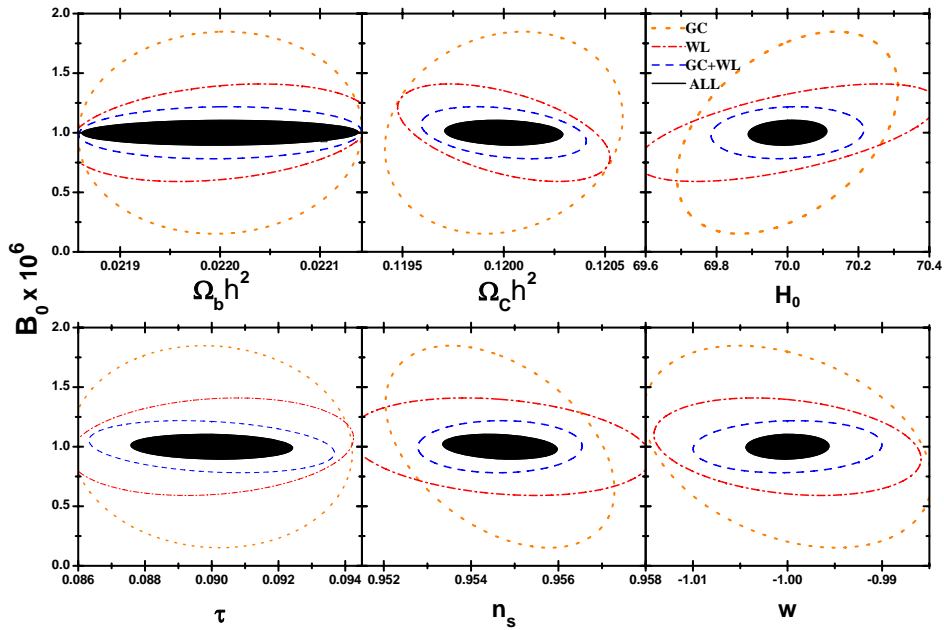}
\caption{The 68\% confidence level contours for the basic parameters and $B_0$. CMB+SNe data is included in all the cases.}
\label{fig:cont-B0}
\end{figure*}

${\Omega_b h^2}$ plays an important role at the recombination, so that the height and separation of the CMB peaks are very sensitive to it. Hence, ${ \Omega_b h^2}$ is well constrained by the CMB data and is not highly correlated with $B_0$. The correlation is significant only when the WL data is considered without the GC data. In this case the constraints on all parameters, including ${ \Omega_b h^2}$, are relatively tight but there is degeneracy between $B_0$ and ${ \Omega_b h^2}$, ${ \Omega_c h^2}$ and $H_0$. Such degeneracies are broken when GC data is included (Figs.~\ref{fig:LSST_B0_wCDM}-\ref{fig:cont-B0}).
The CDM energy density is important after the matter-radiation equality epoch  \footnote{$\Omega_c$ is not directly measured and is inferred from the measuements of total matter or baryonic matter densities.}. CMB data is mainly sensitive to ${ \Omega_c h^2}$ through the geometry of the universe. The SNe Ia data is also sensitive to it through the value of the Hubble parameter entering the distance measurements. Varying ${ \Omega_c h^2}$ will change the epochs of matter-radiation and matter-dark energy equalities, leading to a different history of growth of perturbations. The CDM density is also important for the secondary CMB effects, such as the ISW effect and is an extra source of correlations between the MG parameter, $B_0$, and ${ \Omega_c h^2}$. 
$H_0$ enters in both the distance measurements and the  
evolution equations of matter perturbations, \reff\ref{matter-sub-evolution}). As a result, the MG parameters are highly degenerate with $H_0$ after including the large scale structure data. With the cross-correlation data, the background evolution is discriminated from the matter density perturbations and the degeneracy with $H_0$ is significantly reduced.

The power spectrum of the primordial perturbations is related to $n_s$ and $A_s$ parameters as
\be 
P_{\Psi}(k) \propto A_s k^{n_s -4}\,,
\ee
so that they determine the initial values of scalar metric perturbations, as well as the matter density perturbations. The impact of a modification to growth after recombination can, in principle, be degenerate with varying $n_s$ and $A_s$. The main impact of the optical depth, $\tau$ (not to be confused with conformal time), is on the CMB, including the ISW term, which can be degenerate with evolution of the density perturbations. With the GC and WL data, this degeneracy still exists while the constraints on the parameters become tighter. We therefore see higher correlation in Fig.~\ref{fig:LSST_B0_wCDM} (left). $\tau$ is also correlated with $n_s$ so that tight constraints on $n_s$ can improve errors on $\tau$. Again, with the cross-correlation data, the time evolution of perturbations is much better constrained and the degeneracy is broken. 

The EoS of DE, $w$, like other background parameters, affects the growth by changing the expansion history. The dark energy density, and consequently $w$, influences the recent expansion history and determines the ISW signal. However, since most of the information about $w$ is coming from SNe Ia, the main role of the large scale structure data is to break degeneracies with the MG parameters (Fig.~\ref{fig:cont-B0}).

As can be seen from Fig.~\ref{fig:LSST_B0_wCDM} and later, the bias parameters would have the highest degeneracies with the MG parameters. Observations of galaxies (baryonic matter) are related to the matter density perturbations via the bias parameter. GC data alone is neither able to constrain the bias parameters, nor it can break their degeneracies with $B_0$. \reff\ref{matter_growth-GC}) shows that increasing $B_0$ 
and decreasing the bias have the same impact on the growth of perturbations. 
This negative correlation, however, would vanish at high redshifts due to less observable galaxies. When WL data is added, the sensitivity to the bias parameters, specially at high redshifts, is increased. However, if we look at the uncorrelated plots, we see that the constraints on parameters are shrunk by almost two orders of magnitude. Hence, the residual degeneracy, although relatively high, are not as important. The cross-correlation data is able to lower the uncertainties and the correlations by a factor of few.

While the above discussion provides insight about how our model parameters could be correlated with the MG parameters, with the 1-parameter MG model we can mainly study the theories where there is an intrinsic transition scale in the behavior of gravity due to a scalar degree of freedom with a fixed (f(R)) coupling to matter. In the following subsections, we investigate the degeneracies for models with more MG parameters.

\subsection{Scaler-tensor models parametrized with ($c,s,n$)}
\label{csn}

Fig.~\ref{fig:LSST_csn_wCDM} shows the correlation plots for $c$ (top row), $s$ (middle row) and $n$ (bottom row). Here, $c$ ($\#$7) is a pre-factor controlling the overal amplitude of MG effects, while $s$ ($\#$8) and $n$ ($\#$9) control the redshift and scale of the MG effects, respectively. This parametrization is more flexible in capturing a MG feature, compared to the 1-parameter model in the previous section. However, the MG parameters themselves are highly correlated. The high correlations limit how well the MG parameters can be constrained, and one sees that adding the WL data does not improve the constraints significantly.

From the left panels of Fig.~\ref{fig:LSST_csn_wCDM}, we see that $c$ has a negative correlation with both $s$ and $n$. An increase of $c$ will enhance the growth on all scales. Increasing $n$ decreases the enhancement of growth on larger scales and pushes the MG effects to smaller scales. Hence, the impact of $c$ on the linear growth is opposite to that of $n$. Increasing $s$ pushes the MG effects to more recent times (smaller redshifts). Data is able to discriminate between $c$ and $s$ from the measurement of the growth at multiple redshift bins but $c$ and $n$ are producing highly degenerate effects. $s$ and $n$, on the other hand, are positively correlated because increasing $s$ pushes the MG effects to smaller redshifts. For the WL data, as the dominant observable, this effect is degenerate with increasing $n$, which pushes the MG effects to smaller scales. This degeneracy is an intrinsic property of the WL potential, as described in \cite{Hojjati:2011xd}.

Fig.~\ref{fig:LSST_csn_wCDM} shows that degeneracies of the MG parameters with other parameters are broken significantly after including the WL data. In particular, degeneracies with the bias parameters are reduced so that, $\overline{C}_{i j} = C_{i j}/\sqrt{(C_{ii }\times C_{jj})} \lesssim 0.1$ in all cases. One can see that there is a sweet spot in redshift where the bias parameters are best constrained. This spot in the middle redshift range is where the experiments are most sensitive to MG features \cite{Hojjati:2011xd}.

As we see, adding new MG parameters provides more flexibility in detecting MG effects but the MG parameters will be highly correlated. One suspects that adding even more MG parameters to the model might just weaken the constraints, without providing any new information. In the next section, we consider a model with more MG parameters and test how redundant the extra parameters would be.

\subsection{General model parametrized with BZ parameters}
\label{section:full-BZ}

\begin{figure*}[t]
\begin{center}

\includegraphics[width=2\columnwidth]{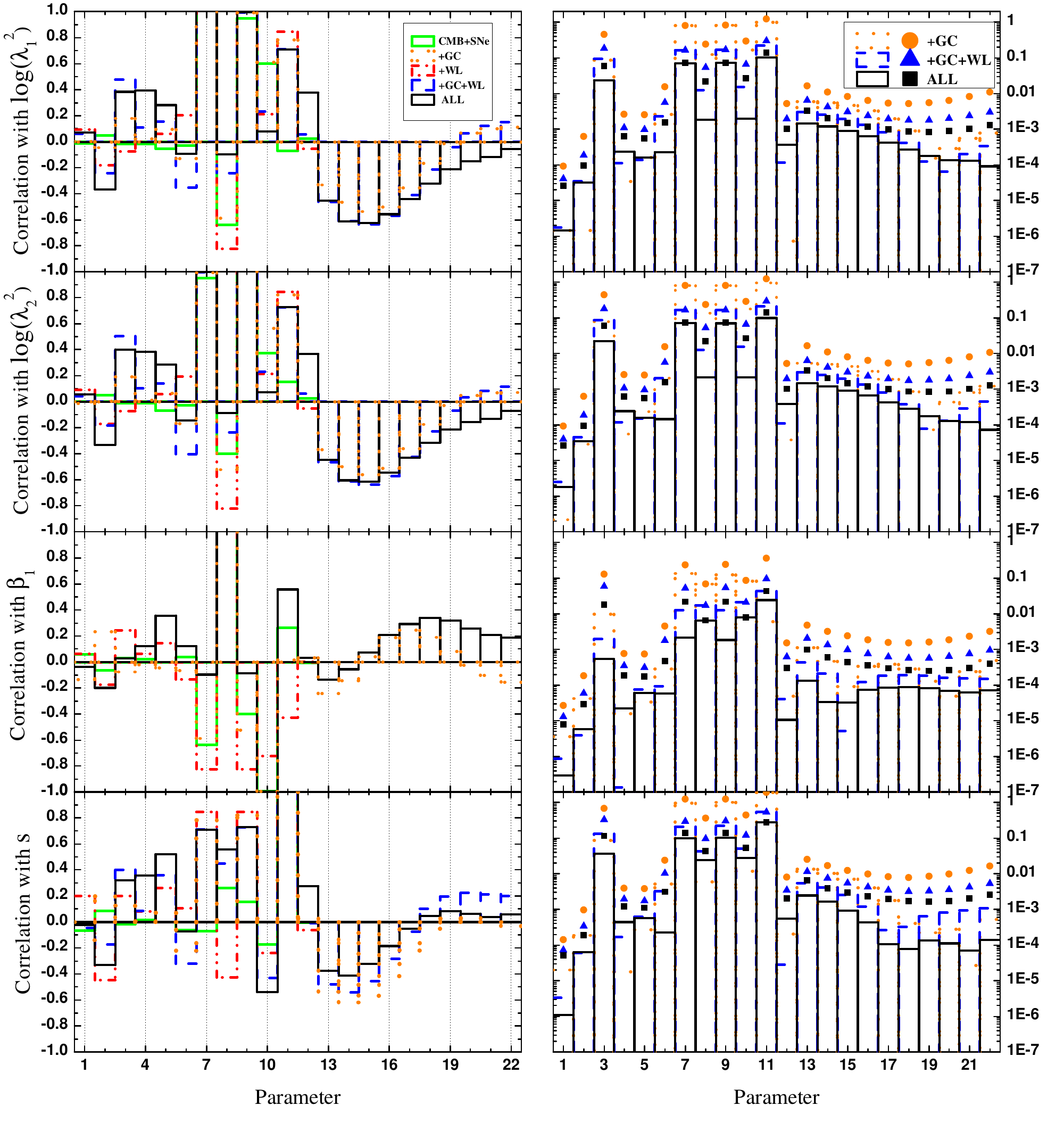}
\end{center}
\caption{Correlation between $\log(\lambda_1^2)$ (first row), $\log(\lambda_2^2)$ (second row), $\beta_1$ (third row) and  $s$ (last row) and other parameters for different combinations of data.}
\label{fig:LSST_BZ_wCDM}
\end{figure*}

Fig.~\ref{fig:LSST_BZ_wCDM} shows the correlations for $\log(\lambda_1^2)$ (first row), $\log(\lambda_2^2)$ (second row), $\beta_1$ (third row) and  $s$ (last row). Here, ($\beta_1$,$\beta_2$) are the analogues of $c$ in the ($c,s,n$) model, while ($\lambda_1$,$\lambda_2$) are similar to $n$, i.e. controlling the transition scale where the MG effects become important. $s$ in both models has the same role. 

The main reason for including this model is to show the redundancy of the extra MG parameters introduced. Comparing the plots in the Figs.~\ref{fig:LSST_csn_wCDM} and \ref{fig:LSST_BZ_wCDM} shows that there are more degeneracies in this model, specially with the bias parameters. We notice that $\lambda_1$ and $\lambda_2$ are constrained very similarly by data. It has been shown that $\mu$ (and hence, $\lambda_1$) should be better constrained than $\gamma$ (hence, $\lambda_2$) by the observables we use here \cite{Hojjati:2011xd}.
But, we see in Fig.~\ref{fig:LSST_BZ_wCDM} that there is almost a perfect correlation between $\lambda_1$ and $\lambda_2$. This correlation limits how well data constrains $\lambda_1$, via the uncertainties of $\lambda_2$ measurement. $\lambda$'s also have a high positive correlation with $s$ ($\#$11) due to the same reason that explained the correlation between $n$ and $s$ in the ($c,s,n$) model. Similarly, $\beta_1$ and $\beta_2$ are also highly correlated and we did not include the correlation plots for $\beta_2$. 

As before, larger $s$ would lead to the enhancement of the growth starting at a larger redshift and accumulating over time. This mimics the effect of a stronger fifth force due to an increase of $\beta$'s. 
Since $s$ is a common parameter to both the BZ model of this section and the ($c,s,n$) model of the previous section, with the same fiducial value, we can get an estimate of how much the constraints on the parameters are diluted and how much extra degeneracy is introduced. Comparing the un-normalized plots for $s$ in Figs.~\ref{fig:LSST_csn_wCDM} and \ref{fig:LSST_BZ_wCDM} confirms that both the errors and the correlations are higher by at least one order of magnitude for the BZ parametrization.

\subsection{Pixelated MG functions}
\begin{figure*}
\centering
\includegraphics[width=2\columnwidth]{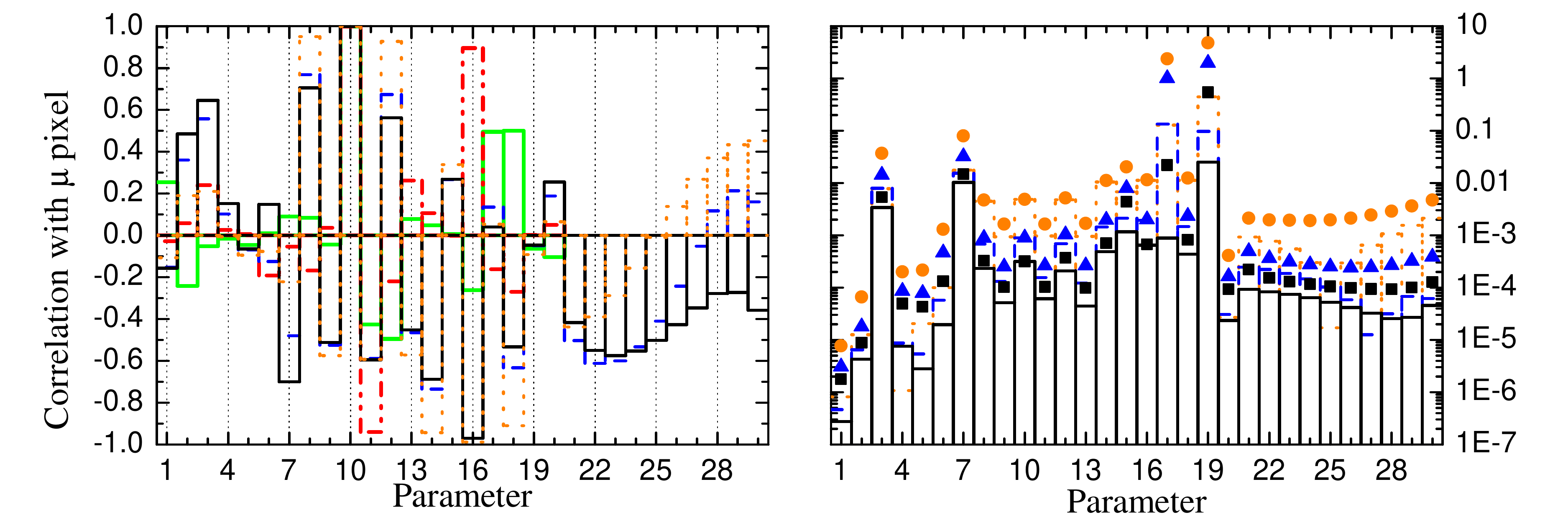}
\includegraphics[width=2\columnwidth]{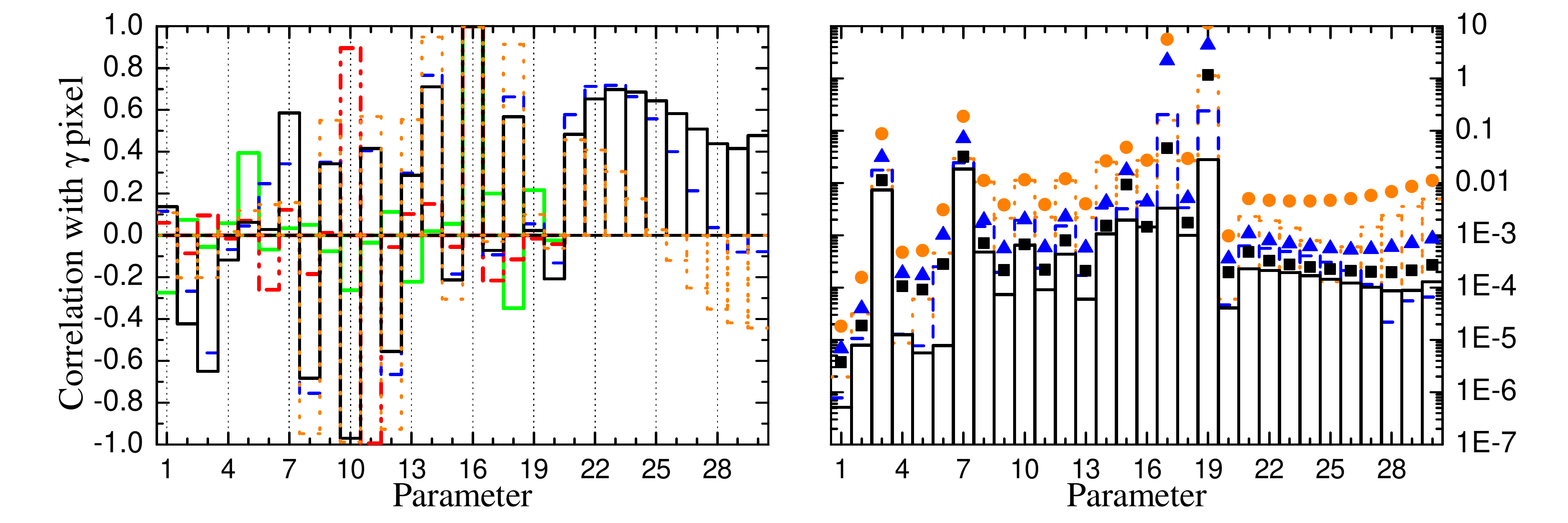}
\caption{Correlation between pixelated and other parameters for different combinations of data.}
\label{fig:LSST_6-6-2-corr}
\end{figure*}

In Fig.~\ref{fig:LSST_6-6-2-corr}, the correlation plots are shown for the middle-k ($0.07 < k < 0.14$ h/Mpc), low-z pixels of $\mu$ (top row) and $\gamma$ (bottom row). We choose these pixels since they cover the scales and redshifts where LSST is most sensitive to the modified growth \cite{Hojjati:2011xd}.

In the previous models, the MG parameters were introduced in a way that they would affect the growth at all scales and redshifts. This inevitably produces correlations between the MG parameters. On the other hand, with the MG pixels, the modified growth is confined to the ranges covered by each pixel so that correlations are intrinsically lower. In particular, the $\mu$ pixels are less correlated with the $\gamma$ pixels compared with the correlations in the BZ model. As a result, the errors on the $\mu$ pixels ($\#$8-13) are obviously smaller than (and are not limited by) those of the $\gamma$ pixels ($\#$14-19). 

The other advantage of working with the MG pixels is that one can isolate some of the physical processes that affect the growth at particular scales or redshifts. For example, we see from Fig.~\ref{fig:LSST_6-6-2-corr} that there is a relatively smaller error on the first high-z $\gamma$ pixel ($\#$ 15) compared to the error on the other two pixels ($\#$ 17,13). This is due to the ISW effect which strongly constrains the $\gamma$ function at large scales \cite{Hojjati:2011xd} which is covered by the first high-z $\gamma$ pixel.

One can also look at the correlations between the MG pixels and $w$ bins (Fig.~\ref{fig:LSST_6-6-2-corr}). There is a high correlation between the MG pixels and the high-z $w$ bin ($\#$7), which is positive (negative) for $\mu$ ($\gamma$). This is reasonable as increasing $w$ would change the onset of the acceleration phase to a later time and hence, more time for the growth of perturbations. On the other hand, the low-z $w$ bin ($\#$6) has a negligible degeneracy with the MG pixels, again, in accordance with what we expect. Since most of the low-z information on $w$ is coming from SNe Ia data and, the low-z $w$ bins are not correlated with the MG pixels.

The plots also show a general trend of negative correlations between the high-z pixels with their corresponding low-z pixels. It makes sense as any change in the high-z (low-z) pixel value should be counteracted with a change in the corresponding low-z (high-z) pixel values.
As for MG parameters in the previous models, there are also high correlations between pixels and basic parameters, like $H_0$ or $A_s$ ($\#$20), and the bias parameters.

\section{Other degenerate effects}
\label{additional}

In this subsection, we take the 3-parameter ($c,s,n$) model and generalize our analysis by including the WL systematics and considering neutrinos to be massive. We would like to know how these additional effects change the correlations between the MG and other parameters. We first include the WL systematics and, in the next step, the massive neutrinos. 

The systematics we consider here are the photo-$z$ errors and some of the errors in the measurement of the point spread function (PSF). These errors are modeled in~\cite{Huterer:2005ez,Zhan:2008jh} and we use their parametrization for our Fisher analysis (see~\cite{Hojjati:2011xd} for details).

Massive neutrinos are taken here as a part of the dark matter (DM) in the universe and their total mass is considered as a parameter in the Fisher analysis, with a fiducial value of $\Sigma {m_{\nu}} = 0.05$ eV.

\begin{figure*}[tbp]
\centering
\includegraphics[width=2.\columnwidth]{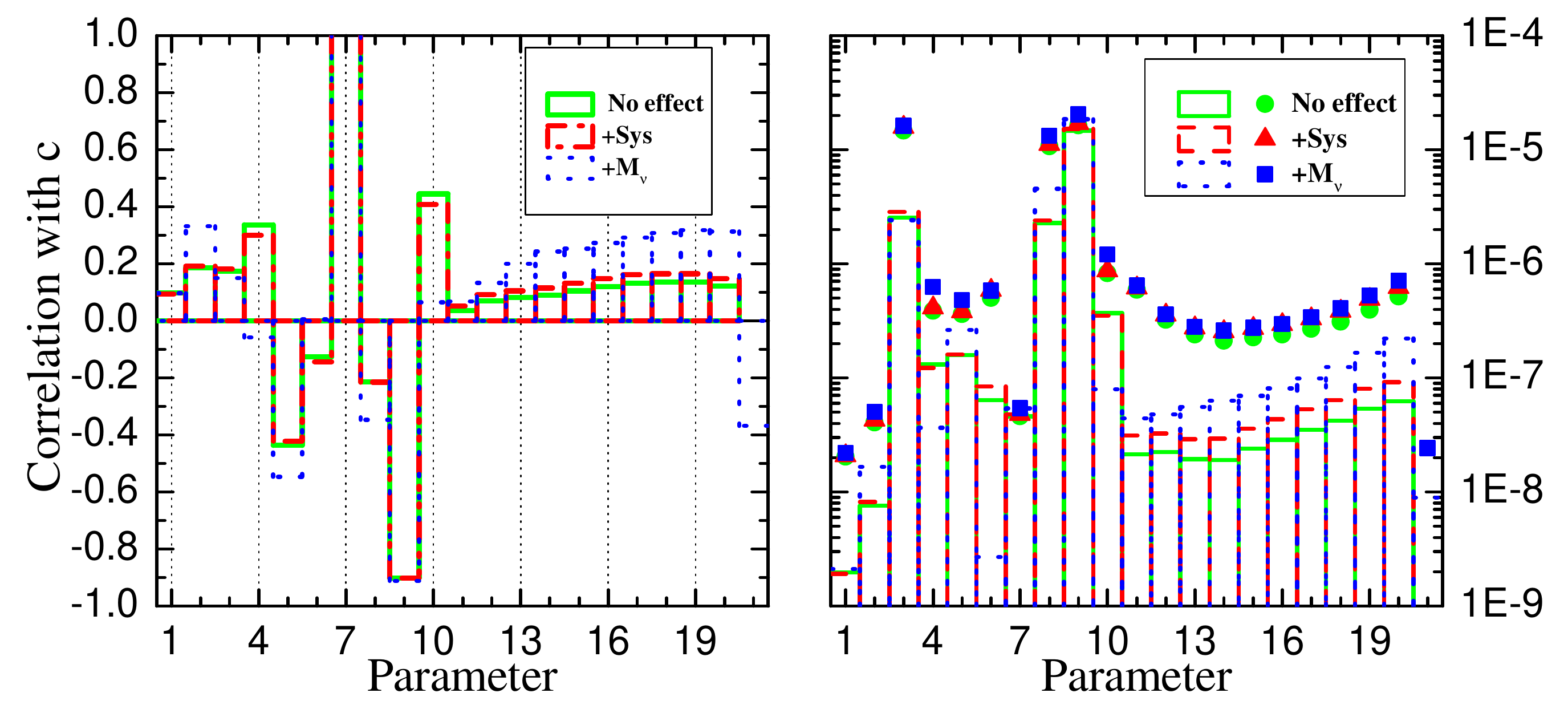}
\includegraphics[width=2.\columnwidth]{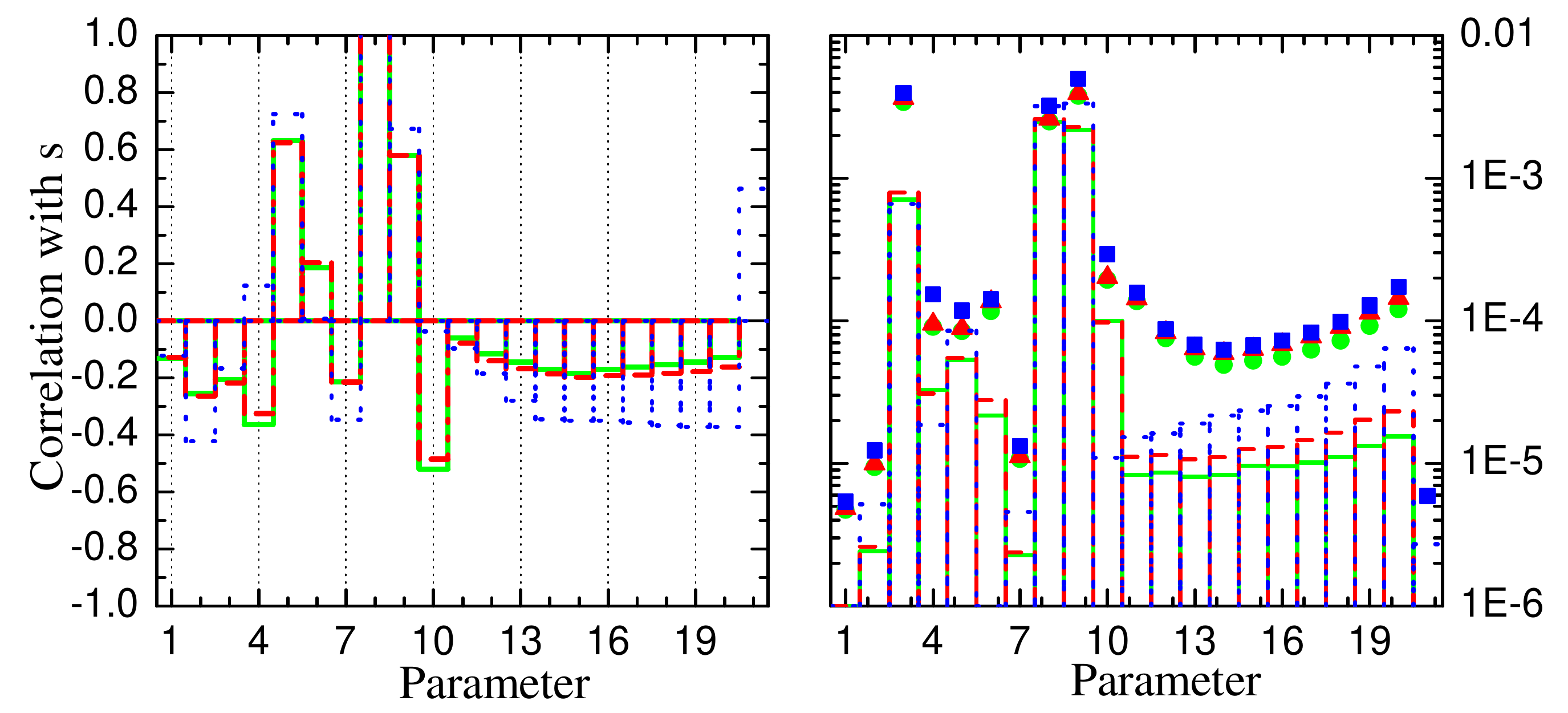}
\includegraphics[width=2.\columnwidth]{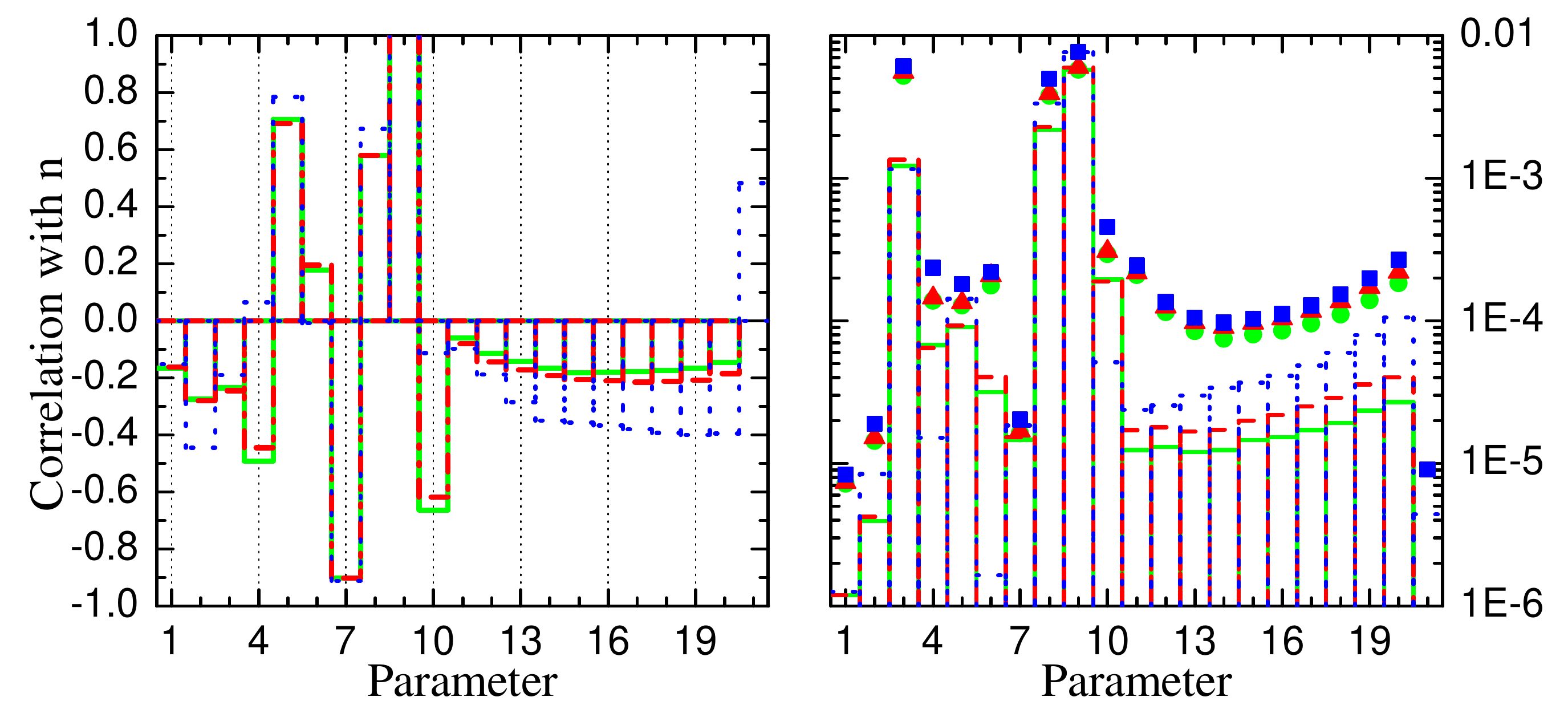}
\caption{correlations for $c$ (top row), $s$ (middle row) and $n$ (bottom row) with full data. Three cases are compared: when no additional effects are included, when the WL systematics included and when neutrinos are massive.}
\label{fig:additional}
\end{figure*}

Fig.~\ref{fig:additional} shows the correlations for $c$ (top row), $s$ (middle row) and $n$ (bottom row) in four cases: without any additional effects, with systematics, with massive neutrinos and with both systematics and massive neutrinos. For these plots we used the full data combination. The total mass of neutrinos is considered as the last parameter ($\#$21) in the plots.

Systematics do not change the parameter constraints and their correlations considerably. With massive neutrinos, however, the correlations of MG parameters with each other, with the bias parameters and some of the basic parameters increases significantly. Massive neutrinos, as part of the DM, would freely stream and slow down the growth on the free-streaming scales. The heavier they are, the larger portion of DM they make, leaving less CDM in the universe. This slows down the growth and, hence, massive neutrinos are negatively correlated with $c$. Heavier neutrinos would have a shorter free-streaming scale. This is roughly the same as the effect of increasing $s$ or $n$, thus, explaining their positive correlation.

\section{Summary}
\label{summary}

In this paper, we studied the degeneracies between different sets of MG parameters with the other parameters that are usually considered in models of large scale structure formation. We used a combination of the CMB, SNe Ia and large scale structure experiments, performed a Fisher analysis of the parameters and computed their correlation matrix.

We considered three models with parametric forms of the MG functions. We saw in Section \ref{Degeneracy} that in the 3-parameter MG model, degeneracies were broken to a good extent but there were relatively high correlations between the MG parameters themselves. It was also demonstrated that increasing the number of the MG parameters would introduce redundancies and weaken the constraints. Finally, with pixelization of the MG functions, we were able to study the degeneracies at different scales and redshifts. 

Among the model parameters, the bias parameters had the highest correlation with the MG parameters, due to their direct involvement in the equation for the growth of perturbations, \reff\ref{matter-sub-evolution}). It seems that for the 3-parameter model, it was possible to break such degeneracies better without weakening the overall constraints on the parameters. The background parameters were degenerate with MG mainly through the expansion history and its impact on the growth time. With the GC and WL data combined, one probes two different combinations of the scalar potentials. This leads to tight constraints and relatively low degeneracies. Among the observables we used, WL provides most of the information about the growth at low redshifts. The intrinsic degeneracy between the scale and the redshift in the WL kernel limits its ability in probing the time evolution of perturbations. This leads to some residual degeneracies between the MG and the basic parameters.
In such cases, the cross-correlation of WL and GC data at multiple tomographic redshift bins will significantly break the degeneracies. In fact, the cross-correlation power spectra are the most informative observables for probing the growth of structure.

We have also considered other physical effects that could be degenerate with a modification to gravity. The WL systematics that we considered in our analysis did not change the correlations while the presence of massive neutrinos could have an important impact. It would be interesting to investigate how including new type of observables, such as redshift space distortion (RSD) \cite{Song:2010fg,Hamilton:1997zq}, would resolve such degeneracies.

\section{Acknowledgement}

The author would like to thank Levon Pogosian for his guidance and useful conversations. AH also benefited from collaboration with Kazuya Koyama, Alessandra Silvestri, Robert Crittenden and specially, Gong-Bo Zhao who shared his CosmoFish code. This work has been supported by an NSERC Discovery Grant and partly by World Class University grant through the National Research Foundation, Ministry of Education, Science and Technology of Korea.

\end{document}